\documentclass[review]{elsarticle}
\usepackage{amsmath}

\usepackage{mathalfa}
\usepackage{mathrsfs}
\usepackage{mathbbol}

\usepackage{lineno,hyperref}
\usepackage{multicol}
\newcommand{\h}{\hspace*{4ex}}

\usepackage{graphicx}
\usepackage{subfig}

\modulolinenumbers[5]

\journal{Journal of \LaTeX\ Templates}

\newcommand{\bb}{\begin{equation}}
\newcommand{\ee}{\end{equation}}
\newcommand{\bega}{\begin{eqnarray}}
\newcommand{\ega}{\end{eqnarray}}
\newcommand{\begae}{\begin{eqnarray*}}
\newcommand{\egae}{\end{eqnarray*}}
\newcommand{\Ncal}{{\cal N}}

\newcommand{\pa}{\partial}

\newcommand{\ug}{ \; = \; }

\newcommand{\om}{\omega}

\begin{document}

\begin{frontmatter}

%\title{Propagation of the frozen waves through stratified media:
%controlling of the longitudinal intensity pattern and characterization
%of antireflective thin films}

\title{Modelling the longitudinal intensity pattern of diffraction resistant beams in stratified media}

%\tnotetext[mytitlenote]{Fully documented templates are available in
%the elsarticle package on \href{http://www.ctan.org/tex-archive/macros/latex/contrib/elsarticle}{CTAN}.}

\author{Grazielle de A. Louren\c{c}o-Vittorino and Michel Zamboni-Rached}
\address{University of Campinas, Campinas, S\~ao Paulo, Brazil}
%\fntext[graz]{email: grazi.lourenzo@gmail.com}
%\fntext[mzr]{email:
% mzamboni@decom.fee.unicamp.br}


\begin{abstract}
In this paper, we study the propagation of the Frozen Wave type
beams through non-absorbing stratified media and develop a
theoretical method capable to provide the desired spatially shaped
diffraction resistant beam in the last material medium. In this context, we also develop a matrix method to deal
with stratified media with large number of layers. Additionally, we undertake some discussion about minimizing reflection of the incident FW beam on the first material interface by using thin films. Our results show that it
is indeed possible to obtain the control, on demand, of the
longitudinal intensity pattern of a diffraction resistant beam
even after it undergoes multiple reflections and transmissions at the layer
interfaces. Remote sensing, medical and military applications, noninvasive optical measurements, etc., are some fields that can be benefited by the method here proposed.

\end{abstract}


\end{frontmatter}

%\linenumbers
\section{Introduction}

\h The propagation of electromagnetic waves through inhomogeneous
media has been widely studied since a long, mainly because it is a
more realistic configuration of what occurs in nature. In general,
the theoretical models dealing with this subject are developed for
plane waves
\cite{jacobssonlight}\cite{lekner1994light}\cite{novitsky2005vector}\cite{Bahaa}\cite{Wait}
and, although researches on Localized Waves (LWs)
\cite{Durnin1987}\cite{Localized-Waves08}\cite{Localized-Waves13}
have proved to be quite advanced, few works have been addressed on
the propagation of such beams along stratified media, often
considering, when it is the case, just a single Bessel beam
\cite{Mugnai2009a}.

\h In the context of the LWs, a very interesting kind of
diffraction-attenuation resistant wave is the so called Frozen
Wave (FW), which possesses great potential for applications wherever
one wishes to shape, on demand, the longitudinal intensity pattern
of electromagnetic wave beams.

\h The theory of FWs was developed in
\cite{Zamboni-Rached2004}\cite{Zamboni-Rached2006} and the
experimental generation of such beams was reported in
\cite{Vieira2012}\cite{Vieira2015}\cite{DORRAH-2016}. Such waves
are constructed by discrete superposition of co-propagating and
equal frequency Bessel beams with different complex amplitudes and
longitudinal wavenumbers, possessing, as the main characteristic,
the possibility of assuming (approximately) any desired
longitudinal intensity pattern within a spatial interval on the
propagation axis, while maintaining the diffraction resistant
properties of the ordinary Bessel beams
\cite{Zamboni-Rached2006}\cite{Zamboni-Rached2009}\cite{Zamboni-Rached2010}.

\h In this paper, we study the behavior of a FW-type-beam when
propagating through non-absorbing stratified media along $z$
direction, showing how the desired beam pattern in the last medium
is affected by the inhomogeneities faced by the beam along its
propagation. To overcome this issue, we develop a novel method for
constructing an incident FW beam capable of compensating such
inhomogeneity effects, so rendering the desired
spatially-shaped-diffraction-resistant-beam in the last material
medium. Additionally, we make some considerations about minimizing the refection of the incident FW beam by using thin films.
Finally, in the last section, we develop a matrix-transfer method to deal
with stratified media with a large number of layers.

\h Our theoretical method works under a scalar model, thus, we are
assuming the paraxial regime, with a linearly polarized wave beam.

\section{A brief overview about the FW method for homogeneous media}

\h A FW beam is given by a superposition of $2N+1$ co-propaganting
and equal frequency Bessel beams \cite{Zamboni-Rached2004} (the
time harmonic term, $e^{-i\omega t}$, is omitted along the entire
paper):

 \bb
\Psi(\rho,\phi,z)= \Ncal_{\nu}
e^{i\nu\phi}\sum\limits_{q=-N}^{N}A_qJ_\nu(h_q\rho)e^{i\beta_qz}
\label{Eq1} \ee
 where
  \bb
h_{q}^{2}=n^2\frac{\om^2}{c^2}-\beta_{q}^2 \,\,\, , \label{Eq2}\ee

$n$ is the refractive index of the medium,
$\om$ the angular frequency, $c$ the light velocity and
$\Ncal_{\nu} =1/[J_{\nu}(.)]_{max}$, with $[J_{\nu}(.)]_{max}$
denoting the maximum value of the $\nu$-order Bessel function of
first kind. The constant coefficients, $A_q$, provide the complex
amplitudes for each Bessel beam in the superposition, while
$h_{q}$ and $\beta_q$ are the Bessel beams transverse and
longitudinal wavenumbers, respectively.

\h The method's main goal is to find out the values of $\beta_q$,
and $A_q$ in Eq.($\ref{Eq1}$), in order to reproduce,
approximately, within $-L/2\leq z\leq L/2$, and over a cylindrical
surface of radius $\rho_{\nu}$, a desired longitudinal intensity
pattern, $|F(z)|^2$, chosen a priori. That is, we wish
$\left|\Psi(\rho=\rho_{\nu},\phi,z)\right|^2\approx|F(z)|^2$. To
obtain such result, we make the following choices
\cite{Zamboni-Rached2004, Zamboni-Rached2006, Zamboni-Rached2005}:

\bb \beta_{q} = Q + \frac{2\pi q}{L} \,\, , \,\,\;\; {\rm with}\,\,\,
q = -N,-N+1,..,N-1,N \label{betaq} \ee and \bb
A_{q}=\frac{1}{L}\int_{-L/2}^{L/2} F(z)e^{-i \frac{2\pi q}{L}}dz \,\, ,
\label{Aq} \ee with $0 \leq \beta_{q} \leq n \om/c  $ and where the value of $Q=a n \om/c$ $(0<a<1)$ can be
freely chosen, since relation (\ref{Eq2}) is obeyed.

\h It is also possible to get some control on the transverse beam
profile, more specifically, we can choose the desired spot radius, $r_0$,
of the resulting beam from the parameter $Q$, via the relation $r_0
\approx 2.4/h_0$, for superpositions of Bessel beam of
zeroth order ($\nu=0$). For $\nu>0$, i.e. higher order FWs, the longitudinal intensity
pattern will be concentrated over a cylindrical surface of radius
$\rho_{\nu}$, whose value corresponds the first positive root of
$[(d/d\rho) J_\nu(\rho h_0)]\mid_{\rho =\rho_{\nu}}=0$.

\h The details of the method can be found in
\cite{Zamboni-Rached2004, Zamboni-Rached2005}.

\section{Propagation of a frozen wave in a nonabsorbing stratified medium}

\h Following a scalar model, Eq.$(\ref{Eq1})$ represents the
transverse cartesian component of the electric field, with a
neglegible longitudinal component (paraxial approximation). In the
case of Bessel beams, such assumption can be made when $h \ll n\om/c$,
resulting in a transverse spot size much larger than the
wavelength.

\textbf{Reflection and transmission}

\h Let us consider a stratified medium formed by $M$ layers with
refractive indexes $n_m$ $(m = 1,2,..,M)$ and with their
interfaces located at the positions $z = d_1,d_2,..,d_{M-1}$. See
Figure 1.

\

\begin{figure}[!h]
\begin{center}
 \scalebox{.55}{\includegraphics{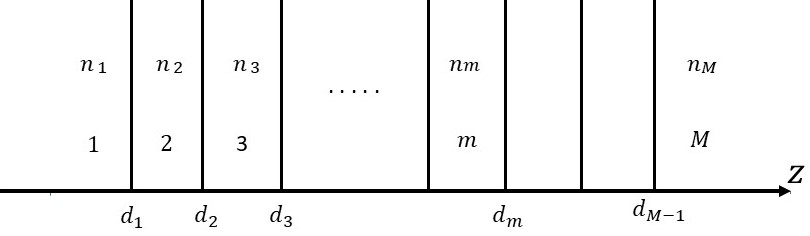}}
\end{center}
\caption{Schematic representation of the stratified medium.}
\end{figure}

\h When a FW (which is a superposition of Bessel beams) coming
from the first medium impinges normally upon the first interface,
a process of multiple reflections and transmissions takes place at each
interface of the stratified medium. It is well known that for a
single Bessel beam in normal incidence on a plane interface, the
reflected and transmitted waves are also Bessel beams with the
same order and with the same transverse wavenumber of the incident
beam, and, naturally, with amplitudes that depend on the
refractive indexes of both media and also on the cone angle of the
incident Bessel beam \cite{Mugnai2009a}.

\h Based on this, we can calculate the field in all $M$ layers
resulting from the normally incident FW beam, $\Psi_{inc}(\rho,\phi,z)$, coming from the
first medium and given by:

\begin{equation}
\Psi_{inc}(\rho,\phi,z)= \Ncal_{\nu}
e^{i\phi\nu}\sum\limits_{q=-N}^{N} A_q J_\nu(h_{
q}\rho)e^{i\beta_{1q}z} \label{inc}
\end{equation}

with

\begin{equation}
h_{q} = \sqrt{n_1^2\frac{\omega^2}{c^2}-\beta_{1 q}^2} \label{h1}
\end{equation}

\h In the $m$th-layer there are forward Bessel beams, $\Ncal_{\nu}
A_q \tau_{m q} J_\nu(h_{q}\rho)\exp(i\nu \phi)\exp(i\beta_{m
q}z)$, and backward ones, $\Ncal_{\nu} A_q \Gamma_{m q}
J_\nu(h_{q}\rho)\exp(i\nu \phi)\exp(- i\beta_{m q}z)$, which
correspond to the incident Bessel beams $\Ncal_{\nu} A_q
J_\nu(h_{q}\rho)\exp(i\nu \phi)\exp(i\beta_{1 q}z)$, with $q =
-N,-N+1,..,N-1,N$. In these previous equations, $\Gamma_{m q}$ and
$\tau_{m q}$ are the reflection and transmission coefficients at
the $m$th and $(M-1)$th interfaces, respectively, and $\beta_{m q}
= \sqrt{n_m^2\om^2/c^2 - h_q^2}$.

In this way, considering the incident FW, Eq.(\ref{inc}), the
total field within each medium will be:

%%\h We write the sum of all contributions of forward and backward waves as $e^{i\beta_{Mq} z}$ (incident beam and transmitted beams) and $ e^{-i\beta_{Mq} z}$ (reflected beams), respectively. If an incident FW $\Psi(\rho,\phi,z)_{inc}=\Ncal_{\nu} e^{i\nu\phi}\sum\limits_{q=-N}^{N} A_q J_{\nu}(h_q\rho)e^{i\beta_{1q}z}$ impinges normally at plane interface localized at $d_1$, the total electric field in each medium, 1 to $M$, can be written as

\begin{align}
   \Psi_{1}(\rho,\phi,z)&= \Ncal_{\nu}
e^{i\nu\phi}\sum\limits_{q=-N}^{N}[A_q J_\nu(h_{
q}\rho)e^{i\beta_{1q}z}+A_q\Gamma_{1q}J_\nu(h_{
q}\rho)e^{-i\beta_{1q}z}]~~~~~(z\leq d_1) \nonumber \\
&\vdots \nonumber\\
   \Psi_{m}(\rho,\phi,z)&= \Ncal_{\nu}
e^{i\nu\phi}\sum\limits_{q=-N}^{N}[A_q\tau_{mq} J_\nu(h_{
q}\rho)e^{i\beta_{mq}z}+A_q\Gamma_{mq}J_\nu(h_{
q}\rho)e^{-i\beta_{mq}z}]~~~~~(d_{m-1}\leq z \leq d_m) \label{psim} \\
   &\vdots \nonumber\\
\Psi_{M}(\rho,\phi,z)&= \Ncal_{\nu}
e^{i\nu\phi}\sum\limits_{q=-N}^{N}A_q\tau_{Mq}J_\nu(h_{
q}\rho)e^{i\beta_{Mq}z} ~~~~~(z\geq d_{M-1}) \nonumber
\end{align}
with\footnote{Naturally, $\tau_{1 q}=1$ and $\Gamma_{M q}=0$.}

\bb \beta_{m q} = \sqrt{n_m^2\frac{\om^2}{c^2} - h_q^2}  \;\;\;\;\;\;\;\;\;\;\;\;\;  m = 1,2,..,M \label{betamq} \ee

\h In order to evaluate the reflection and transmission
coefficients, $\Gamma_{mq}$ and $\tau_{mq}$, respectively, we must
apply the boundary conditions on each interface. Such conditions
(expressing the continuity of the tangential electric and magnetic
fields) are given by:

\begin{align}
   \Psi_{1}(\rho,\phi,z)\vert_{z=d_1}&=\Psi_{2}(\rho,\phi,z)\vert_{z=d_1} \nonumber  \\
   \frac{\partial \Psi_{1}(\rho,\phi,z)}{\partial z}\vert_{z=d_1}
&=\frac{\partial \Psi_{2}(\rho,\phi,z)}{\partial z}\vert_{z=d_1} \nonumber \\
   &\vdots    \label{cc} \\
   \Psi_{M-1}(\rho,\phi,z)\vert_{z=d_{M-1}}&=\Psi_{\tau M
}(\rho,\phi,z)\vert_{z=d_{M-1}} \nonumber \\
\frac{\partial \Psi_{M-1}(\rho,\phi,z)}{\partial
z}\vert_{z=d_{M-1}}&=\frac{\partial \Psi_{\tau
M}(\rho,\phi,z)}{\partial z}\vert_{z=d_{M-1}} \nonumber
\end{align}

\h By using Eqs.(\ref{psim}) in Eqs.(\ref{cc}), we can find out all
$\Gamma_{mq}$ and $\tau_{mq}$ and, therefore, fully characterize
the total field in each layer through Eqs.(\ref{psim}).

\h In the next two examples, we will use the equations developed
in this section to precisely describe the propagation of a FW
through stratified media with few layers. More specifically, we
will see very clearly how the desired intensity pattern of a FW,
\emph{initially designed for an homogeneous medium}, is affected after
passing through all the material layers, suffering multiple
reflections and transmissions.

\h In all examples of this paper we will use the angular frequency $\om = 2.9788 \times 10^{15}$rad/s, which correspond to $\lambda = 632.8$nm in vacuum.

\textbf{First example}

\h Let us consider a stratified structure formed by four different
media with refractive indexes $n_1 = 1$, $n_2 = 3$, $n_3 = 2.5$,
$n_4 = 4$ and whose interfaces are located at $d_1=0$, $d_2=0.01$m
and $d_3=0.025$m. Considering an incident beam given by
Eq.(\ref{inc}) with $ \nu = 0$, we know that the field within each layer is given
by Eqs.(\ref{psim}), and by using Eqs.(\ref{cc}) we find the
reflection and transmission coefficients

\begin{equation}
\Gamma_{1q}=\frac{P^+
\beta_{1q}-P^-\beta_{2q}}{P^-\beta_{2q}+P^+\beta_{1q}} \label{Eq8}
\end{equation}
\begin{equation}
\Gamma_{2q}=\frac{\beta_{1q}(1-\Gamma_{1q})}{\beta_{2q}(e^{2i\alpha_{1q}}\xi-1)}
\label{Eq9}
\end{equation}
\begin{equation}
\tau_{2q}=\Gamma_{2q}e^{-2i\alpha_{1q}}\xi \label{Eq10}
\end{equation}
\begin{equation}
\Gamma_{3q}=\frac{T_{2q}e^{i\alpha_{1q}}+\Gamma_{2q}e^{-i\alpha_{1q}}}{e^{-i\alpha_{2q}}-e^{i(\alpha_{2q}-2\alpha_{3q})}\Phi}
\label{Eq11}
\end{equation}
\begin{equation}
\tau_{3q}=-\Gamma_{3q} e^{-2i\alpha_{3q}}\Phi \label{Eq12}
\end{equation}
\begin{equation}
\tau_{4q}=\frac{\beta_{3q}(T_{3q}e^{i\alpha_{3q}}-\Gamma_{3q}e^{-i\alpha_{3q}})}{\beta_{4q}e^{i\alpha_{4q}}}
\label{Eq13}
\end{equation}
with
\begin{equation}
\Phi=\frac{\beta_{4q}+\beta_{3q}}{\beta_{4q}-\beta_{3q}}
\label{Eq14}
\end{equation}
\begin{equation}
\xi=\frac{\beta_{2q}(e^{-i\alpha_{2q}}-e^{i(\alpha_{2q}-2\alpha_{3q})}\Phi-\beta_{3q}(e^{-i\alpha_{2q}}+e^{i(\alpha_{2q}-2\alpha
_{3q})}\Phi)}{\beta_{2q}(e^{-i\alpha_{2q}}-e^{i(\alpha_{2q}-2\alpha_{3q})}\Phi+\beta_{3q}(e^{-i\alpha_{2q}}+e^{i(\alpha_{2q}-2\alpha
_{3q})}\Phi)} \label{Eq15}
\end{equation}

and

\begin{eqnarray*}
P^+=e^{-2i\alpha_{1q}}\xi+1~~~~~~~~~~P^-=e^{-2i\alpha_{1q}}\xi-1
\end{eqnarray*}

where
$\alpha_{1q}=\beta_{2q}d_2~~~~~~\alpha_{2q}=\beta_{3q}d_2~~~~~~\alpha_{3q}=\beta_{3q}d_3~~~~~~\alpha_{4q}=\beta_{4q}d_3$

\h The values of the longitudinal wavenumbers, $\beta_q$, and the
coefficients $A_q$ of the incident FW are still unknown and should depend on the beam spatial structure we wish for the last
medium. But, due to the fact that, for now, we do not have a method to
get the desired diffraction resistant beam in the last medium
(after the stratified structure), we will design the incident beam
as if we were dealing with an homogeneous medium (the first one,
in this case) and, thus, we will see the effects suffered by the
beam after it has crossed the stratified structure.

\h Now, let us suppose we wish, in the last medium, a diffraction resistant beam with a spot radius
$r_0 \approx 4.5\mu$m and whose desired longitudinal intensity
pattern, $|F(z)|^2$, has a ladder shape of three steps. More
specifically, within $-L/2 \leq z \leq L/2$,

\bb F(z)\left\{ \begin{array}{c}
\sqrt3 ~~~~~~~~l_1\leq z<l_2\\
\sqrt2 ~~~~~~~~l_2\leq z<l_3\\
~1~~~~~~~~~l_3\leq z\leq l_4\\
~0~~~~~~~~~~$elsewhere$
\\
\end{array}
\right.\label{F}\ee where $L=0.30$m, $\delta=0.035$m,
$l_1=0.03$m, $l_2=l_1+\delta$, $l_3=l_2+\delta$ and $l_4=l_3+\delta$. The value of $Q$ is
obtained from the desired spot radius, resulting in
$Q=0.99856n_1\om/c$. In this case, let us adopt the value $N=20$. The coefficients $A_{q}$ are calculated
inserting the function $(\ref{F})$ in $(\ref{Aq})$, and the
longitudinal wavenumbers $\beta_{1q}$ by $(\ref{betaq})$, being the
values of $h_q$ given by Eq.(\ref{h1}).
%\begin{eqnarray*}
%h_{q}=\sqrt{(n_1k)^2 - (\beta_{1q})^2 }
%\label{}\end{eqnarray*}

\h By knowing the values of $h_{q}$, we can determine the sets of
longitudinal wavenumbers in the second, third and fourth medium
through $\beta_{m q} = \sqrt{n_m^2\om^2/c^2 - h_q^2}$, $m=2,3,4$.
Thus, we can write the transmitted field in the medium 4 as
\begin{equation}
\Psi_{4}(\rho,\phi,z)=
\Ncal_{\nu}\sum\limits_{q=-N}^{N}A_q\tau_{4q}J_0(h_{
q}\rho)e^{i\beta_{4q}z} ~~~~~(z\geq d_{3}) \label{Eq16}
\end{equation}
% trans_sc_3d
\begin{figure}[!htb]
\centering
\subfloat[]{
\includegraphics[height=5.5cm]{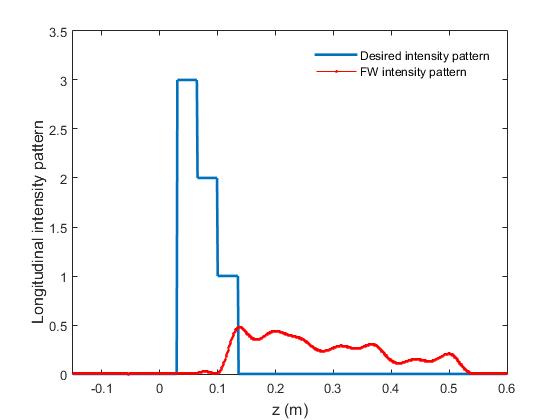}
\label{}
}
%\quad %espaco separador
\subfloat[]{
\includegraphics[height=6.2cm]{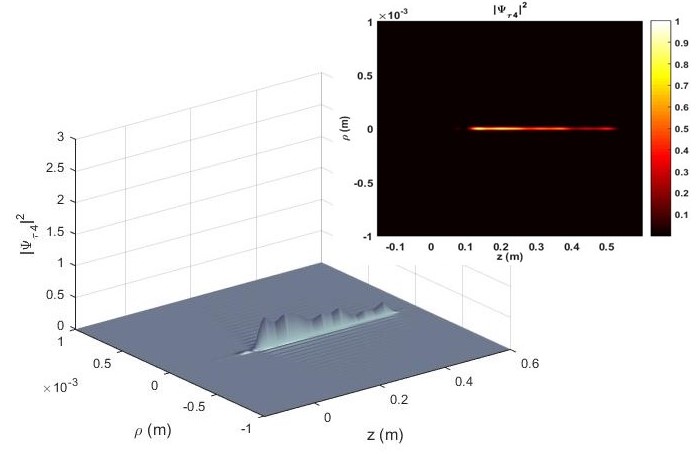}
\label{}
}
\caption{(a) Comparison between the on-axis longitudinal intensity of the FW (with $\nu=0$) and desired intensity pattern, $|F(z)|^2$; (b) the 3D field intensity of the resulting beam in medium 4, as well as its orthogonal projection in the detail.}
\label{fig01}
\end{figure}

\h Figure 2(a) shows the desired longitudinal intensity pattern, $|F(z)|^2$, and the on-axis intensity of the resulting field through the four media. We can see that the characteristics of the FW beam are very much affected when compared to the desired pattern. Figure 2(b) show the 3D field intensity of the resulting wave, again in all media, with its orthogonal projection in the detail.

\textbf{Second example}

\h In this example we consider a stratified medium of four layers, with refractive indexes (as before) $n_1 = 1$, $n_2 = 3$, $n_3 = 2.5$ and $n_4 = 4$, whose interfaces are located at $d_1=0$, $d_2=5$mm
and $d_3=10$mm. At this time we are going to deal with a hollow FW so, in the incident beam solution, Eq.(\ref{inc}), we choose $\nu=4$. The resulting field within each layer is given by Eq.(\ref{psim}). The reflection and transmission coefficients can be found through Eqs.(\ref{psim},\ref{cc}), resulting in Eqs.(\ref{Eq8})-(\ref{Eq13}).

\h Here, we wish, in the last medium, a diffraction resistant beam concentrated on a cylindrical surface of radius $\rho_4 \approx 3\mu$m and with a longitudinal intensity pattern, $|F(z)|^2$, of a ladder shape of three steps, according to Eq.(\ref{F}), but now with $L=50$ mm, $\delta=L/100$, $l_1=(L/4+4\delta)$, $l_2=(l_1+4\delta)$, $l_3=(l_2+4\delta)$ and $l_4=(l_3+4\delta)$. Again, as we do not have, for now, a method to
get the desired diffraction resistant beam in the last medium, we design the incident beam
as if we were dealing with an homogeneous medium (the first one). So, from the desired cylindrical surface radius we obtain $Q=0.9839n_1\om/c$ and from Eq.(\ref{Aq}) we calculate the coefficients $A_q$. The longitudinal wavenumbers are obtained from Eq.(\ref{betaq}).

\h The field within each layer is given by Eqs.(\ref{psim}), from where we get for the last (fourth) medium:

\begin{equation}
\Psi_{4}(\rho,\phi,z)=\Ncal_{\nu} e^{4i\phi}
\sum\limits_{q=-N}^{N}A_q\tau_{4q}J_4(h_{ q}\rho)e^{i\beta_{4q}z}
~~~~~(z\geq d_{3}) \label{Eq17}
\end{equation}

\h Figure 3(a) shows the desired longitudinal intensity pattern, $|F(z)|^2$, to occur on the cylindrical surface of radius $\rho_4 = 3\mu$m, and the corresponding intensity pattern of the resulting beam occurring through the four material media. We can see how the characteristics of the FW beam are affected when compared to the desired pattern. Figure 3(b) show the 3D field intensity of the resulting wave, again in the four media, with its orthogonal projection in the detail.

\begin{figure}[tb]

\subfloat[]{
    \includegraphics[height=5.5cm]{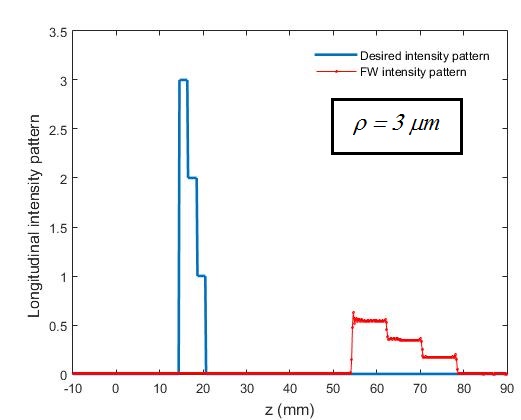}
}
\subfloat[]{
     \includegraphics[height=6.2cm]{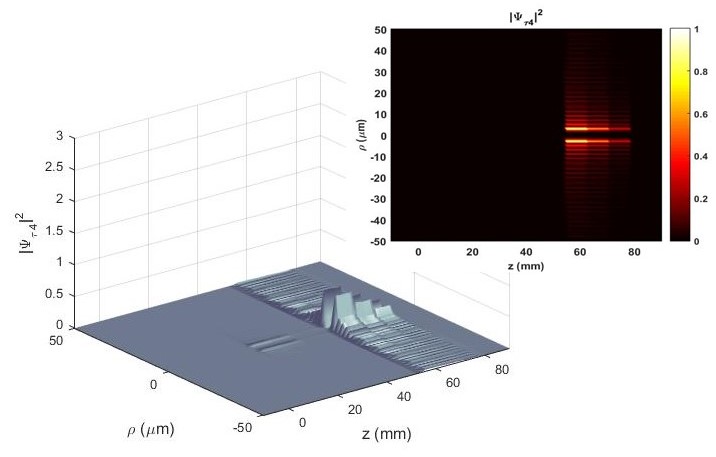}
}
\caption{(a) Comparison between the longitudinal intensity of the FW (with $\nu=4$) over a cylindrical surface of radius $\rho_4 \approx 3\mu$m  and desired intensity pattern $|F(z)|^2$; (b) the 3D field intensity of the resulting beam in medium 4, as well as its orthogonal projection in the detail.}
 \label{gdimotes}
\end{figure}

%figure

\h In general, such deformations inflicted to the FW beam by the stratified structure will occur with more or less intensity, depending on the refractive indexes and thickness of the slabs between the first and the last media.

\section{The compensation method}

\h Our main objective in this work is to develop a method to
obtain a FW beam in the last material medium of a stratified structure, that is, a beam resistant to diffraction effects and whose
longitudinal intensity pattern can be chosen on demand. As we
have seen in the previous sections, a FW designed for an
homogeneous medium can undergo significant changes on its spatial
structure (previously chosen) when crossing a layered medium
before reaching the desired spatial region, which we consider here
as being in the last material medium.

\h In this section, we will
develop what we call the \emph{\emph{compensation method}}, which consists in
designing a beam in the first material medium such
that, when crossing the stratified structure ahead, the changes of
amplitude and phase undergone by its constituent Bessel beams will eventually transform it into the desired FW,
i.e., into the diffraction resistant beam endowed with the desired
longitudinal intensity pattern.

\h The method is presented in four steps described below:

\begin{enumerate}
  \item{First, once it is given the stratified medium, composed of $M$ layers, and considering the incident beam of the type given by Eqs.(\ref{inc},\ref{h1}), where we still do not know the values of the coefficients, $A_q$, a nd the longitudinal wavenumbers, $\beta_{1 q}$, we use Eqs.(\ref{psim},\ref{cc}) to calculate the reflection and transmission coefficients, $\Gamma_{M q}$ and $\tau_{M q}$, respectively, in function of the refractive indexes, $n_M$, the interface positions, $d_M$, and the longitudinal wavenumbers, $\beta_{M q}$ (still unknown).}
  \item{Afterwards, we choose the characteristics of the diffraction resistant beam that we wish to occur in the last medium, $z\geq d_{M-1}$, i.e., we choose the beam spot size, $r_0$, or the radius of the cylindrical surface, $\rho_{\nu}$ (in the case of a hollow beam), and also the desired longitudinal intensity pattern, $|F(z)|^2$, which has to be defined within $-L/2 \leq z \leq L/2$; such spatial range must be large enough to enclose part of the first medium, all the stratified structure and part of the last medium, which must include the region where the desired beam pattern will be located.}
  \item{We directly construct the desired beam solution, $\Psi_M(\rho,\phi,z)$ in the \emph{last} medium through the following equations:

  \bb \Psi_{M}(\rho,\phi,z) = \Ncal_{\nu} e^{i\nu\phi}\sum\limits_{q=-N}^{N}A_q\tau_{Mq}J_\nu(h_{q}\rho)e^{i\beta_{Mq}z} \label{psiM} \ee
with

\bb \beta_{M q} = Q + \frac{2\pi}{L}q \,\,\, , \label{betaMq} \ee

\bb h_q = \sqrt{n_M^2\frac{\om^2}{c^2} - \beta_{M q}^2}  \label{hq2} \ee
and

\bb
A_{q}\tau_{M q} = \frac{1}{L}\int_{-L/2}^{L/2} F(z)e^{-i \frac{2\pi q}{L}}dz
\label{Aqtau} \ee

With $Q$ obtained from $r_0 = 2.4 / \sqrt{n_M^2\om^2/c^2 -
Q^2}$ (with $\nu=0$) or from $[(d/d\rho) J_\nu(\rho
\sqrt{n_M^2\om^2/c^2-Q^2})]\mid_{\rho =\rho_{\nu}}=0$ (with $\nu \geq 1$ of our
choice), where $|F(z)|^2$, $r_0$ and $\rho_{\nu}$ have been chosen
in the previous step.

Note that, now, once we have the values of $h_q$, the reflection
and transmission coefficients, analytically calculated in the
first step, can be numerically evaluated.}

  \item{This final step is directed to calculate the incident beam, $\Psi_{inc}(\rho,\phi,z)$, i.e., the values of $A_q$ and $\beta_{1 q}$ to be used in Eq.(\ref{inc}), in such way that the beam in the last medium will result to be the desired one and already calculated in step 3. Actually, this can be done through Eqs.(\ref{betaMq}-\ref{Aqtau}), from which we get:

      \bb \beta_{1 q} = \sqrt{n_1^2\frac{\om^2}{c^2} - h_q^2} = \sqrt{(n_1^2-n_M^2)\frac{\om^2}{c^2} - (Q + \frac{2\pi}{L}q)^2}  \,\,\, , \label{beta1q2} \ee

      \bb A_{q} = \frac{1}{\tau_{M q}L}\int_{-L/2}^{L/2} F(z)e^{-i \frac{2\pi q}{L}}dz \,\,\, ,
\label{Aq2} \ee
with $h_q$ given by Eq.(\ref{hq2}).

\h Naturally, the resulting field within the $m$th medium can be
calculated via Eqs.(\ref{psim},\ref{betamq}).}

\end{enumerate}

\subsection{The method applied to FWs}

\h Now, we are going to use the compensation method to the
previous examples (1) and (2) for obtaining the desired FW beam in
the last medium of the multilayered structure.

\textbf{First example revisited}

\h Returning to the first example, we note that the first and
second steps of the compensation method were already implemented
through the Eqs.(\ref{Eq8}-\ref{Eq13},\ref{F}) and the chosen spot
radius ($r_0 = 4.5\mu$m). Now, we implement the third step by
constructing the solution for the desired beam in the fourth
medium (the last one) through Eqs.(\ref{psiM}-\ref{Aqtau}), with
$M=4$. In this case, the value of $Q$ results to be $Q = 0.99991
n_4 \om/c$.

\h Finally, we go through step four and calculate the incident
beam via Eq.(\ref{inc},\ref{beta1q2},\ref{Aq2}). By construction,
this incident beam, after crossing the stratified structure ahead,
will result, in the last medium, into the diffraction resistant
beam endowed with the desired longitudinal intensity pattern
(obtained in the previous paragraph). The resulting fields within
the second and third media are calculated via
Eqs.(\ref{psim},\ref{betamq}), with $m=2,3$.

\h Figure 4(a) shows the desired intensity pattern, $|F(z)|^2$,
and the intensity of the resulting field, from the first until the
last medium. We can see a very good agreement between them. Figure
4(b) shows the 3D intensity of the resulting FW beam through the
four media. It is very clear that the compensation method provides
the desired diffraction resistant beam in the last medium. In
this case, we use $N=20$.

\

\textbf{Second example revisited}

\h In applying the compensation method to the second example, we
note that the first and second steps were already implemented
through the Eqs.(\ref{Eq8}-\ref{Eq13},\ref{F}) and the chosen of
the cylindrical surface's radius $\rho_4 = 3\mu$m. The step three,
i.e, the construction of the desired beam solution in medium four,
is implemented through Eqs.(\ref{psiM}-\ref{Aqtau}), with $\nu=4$
and $M=4$. Here, the value of $Q$ results to be $Q =
0.999n_4\om/c$. The fourth step is executed by calculating the
incident beam via Eq.(\ref{inc},\ref{beta1q2},\ref{Aq2}). The
resulting fields within the second and third media are calculated
via Eqs.(\ref{psim},\ref{betamq}), with $m=2,3$. In this case, we
use $N=300$.

\h Figure 5 shows that the resulting FW beam has the very same
spatial shape chosen \emph{a priori}, so confirming the efficiency
of the method.

\begin{figure}[tb]
\subfloat[]{
    \includegraphics[height=5.5cm]{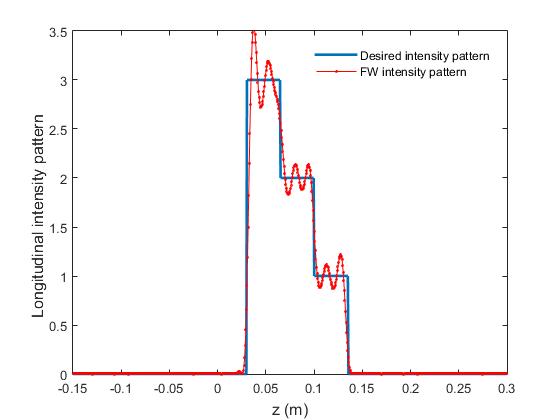}
}
\subfloat[]{
    \includegraphics[height=6.2cm]{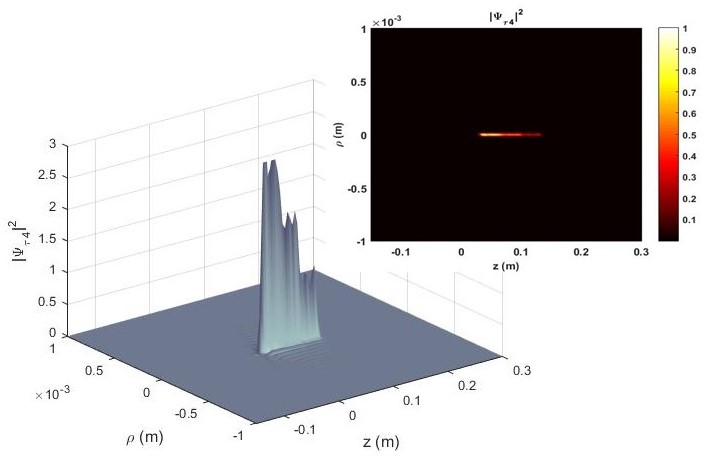}
}
\caption{ The resulting FW (with $\nu=0$) in medium 4 after applied the compensation method. (a) Comparison between the on-axis longitudinal intensity of the FW and desired pattern, $|F(z)|^2$; (b) the three-dimensional field intensity of the resulting FW, as well as its orthogonal projection in the detail.}
 \label{gdimotes}
\end{figure}

\begin{figure}[tb]
\subfloat[]{
    \includegraphics[height=5.5cm]{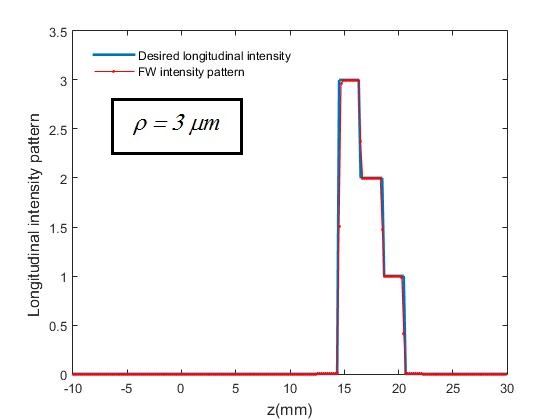}

}
\subfloat[]{
     \includegraphics[height=6.2cm]{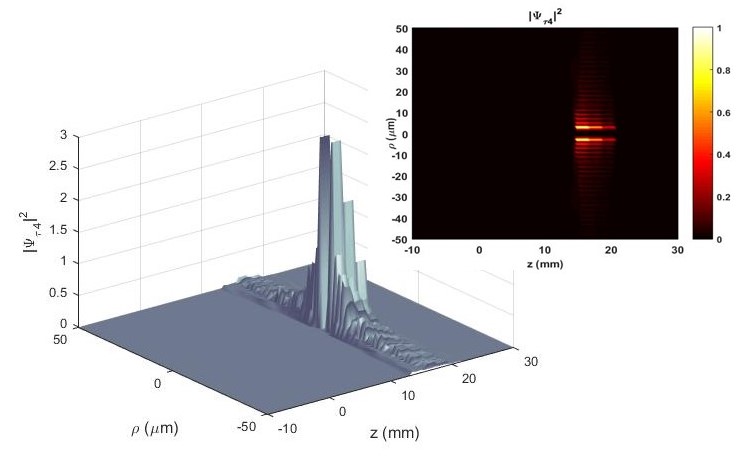}
     }

\caption{The resulting FW (with $\nu=4$) in medium 4 after applying the compensation method. (a) Comparison between the longitudinal intensity of the FW over the cylindrical surface of radius $\rho=\rho_4=3\mu$m and the desired intensity pattern, $|F(z)|^2$; (b) the 3D field intensity of the resulting FW, as well as its orthogonal projection in the detail.} \label{gdimotes}
\end{figure}

\subsection{Restriction over the longitudinal wavenumbers}

\h Let us stress that our preference here is to deal with
propagating waves whenever it is possible. For this, all longitudinal
wavenumbers  have to be real in all layers. In the last medium,
this condition is guaranteed by construction, since the $\beta_{M
q}$ are chosen according to Eq.(\ref{betaMq}). Now, within the
$m$th medium the longitudinal wavenumbers, $\beta_{m q}$ will be
real if

\begin{eqnarray*}
(\beta_{m q})^2 = n_m^2\frac{\om^2}{c^2} - h_{q}^2 = (n_m^2 - n_M^2)\frac{\om^2}{c^2} + \left(Q + \frac{2\pi}{L}q \right)^2 \geq 0 \,\, , \label{cond1}
\end{eqnarray*}
where we have used Eq.(\ref{hq2}).

\h A sufficient (but not necessary) condition for getting
inequation (\ref{cond1}) satisfied for all $m = 1,2,..,M-1$ and
$q=-N,..,0,..,N$ is to have

\begin{equation}
[(\beta_{M q})_{min}]^2 = \left(Q - \frac{2\pi}{L}q \right)^2 \geq (n_M^2 - n_m^2)\frac{\om^2}{c^2}
 \label{cond2}
\end{equation}

\h In case inequation (\ref{cond1}) is not satisfied for one or
more values of $q$ within the $m$th medium, we have to change the
value of the parameter $Q$ and/or the value of $N$, depending on
the case.

\section{Minimizing the reflection}

\h Anti-reflection coating is a very important application of the
optics of thin films
\cite{Southwell-1983}\cite{Lowdermilk-Milam-1980}\cite{Chen-2012}. The technics used for minimizing reflection of ordinary beams
when impinging on dielectric surfaces can also be used when the
incident wave is a FW-type beam. In this section we will analyze
the very simple case of determining the refractive index and the
thickness of a film to be used on the separation plane surface of
two dielectric media for minimizing the reflection of a normally
incident FW beam.

\h The reader may note that we say ``minimize'' rather than
``nullify'' reflection. This is due to the fact that a FW is
composed of a superposition of Bessel beams with different
longitudinal wavenumbers, which makes it impossible to extinguish
the reflection coefficients of all them simultaneously. Thus, we
have to choose the most important Bessel beam of the superposition
and characterize the thin film that extinguishes its reflection.
In doing so, we will minimize the reflection of the incident FW,
since, in general, the other Bessel beams composing it differ
little from the main Bessel beam with respect to the values of the
longitudinal wavenumbers.

\h In general, the Bessel beam that most contribute to the
superposition that defines a FW, like that given by
Eq.(\ref{psiM}), is the central one, which corresponds to $ q = 0
$. The anti-reflective film will be characterized considering this
Bessel beam. In cases where the most important beam of the FW is
not the central one (as it may occur when the function
$F(z)$, which defines the desired longitudinal pattern, oscillates
considerably), one should first find which Bessel beam is the most
relevant to the superposition and perform the characterization of
the thin film based on it.

\h Before to proceed, let us remember that a Bessel beam is
characterized for its order and its cone-angle $\theta$, which
determines its longitudinal and transverse wavenumbers, $\beta = (n
\om/c)\cos\theta$ and $h = (n \om/c)\sin\theta$ \cite{Durnin1987}.
From the compensation method we already know that the central
Bessel beam of the resulting FW in the last medium (the third one,
in this case) has its longitudinal wavenumber $\beta_{3\,0} = Q
\equiv \bar{\beta_3}$. The longitudinal wave numbers of the
correspondent central Bessel beams in media 1 and 2 will be
$\beta_{1\,0} = \sqrt{(n_1^2-n_3^2)\frac{\om^2}{c^2} - Q^2} \equiv
\bar{\beta_1}$ and $\beta_{2\,0} =
\sqrt{(n_2^2-n_3^2)\frac{\om^2}{c^2} - Q^2} \equiv \bar{\beta_2}$,
respectively. The cone angle of these Bessel beams are given by
$\theta_1 = \arccos(c\,\bar{\beta_1}/n_1\om)$, $\theta_2 =
\arccos(c\,\bar{\beta_2}/n_2\om)$ and $\theta_3 =
\arccos(c\,Q/n_3\om)$.

\h Now, considering the system formed by the three parallel layers
(the intermediate layer is the film with thickness $d$ and
refractive index $n_2$), the (central) incident Bessel beam is
totally transmitted if its reflection coefficient is equal to
zero, that is:

  \begin{equation}
\overline{\Gamma}_{1}=\frac{\overline{\beta}_{2}(1+e^{-2i\alpha}\xi)+\overline{\beta}_{1}(1-e^{-2i\alpha}\xi)}{\overline{\beta}_{1}(1-e^{-2i\alpha}\xi)-\overline{\beta}_{2}(1+e^{-2i\alpha}\xi)}=0
\label{Eq27}
\end{equation}
where $\alpha=\overline{\beta}_2d$ and
$\xi=(\overline{\beta}_{3}+\overline{\beta}_{2})/(\overline{\beta}_{3}-\overline{\beta}_{2})$. From Eq.(\ref{Eq27}), we get the following system of two equations

\begin{align} \left\{ \begin{array}{l}
\overline{\beta}_2+\overline{\beta}_2\xi\cos(2\alpha)=-\overline{\beta}_1+\overline{\beta}_1\xi\cos(2\alpha)  \\
\overline{\beta}_2\sin(2\alpha)=\overline{\beta}_1\sin(2\alpha)  \,\,\,, \label{Eq29} \\
\end{array}
\right.\end{align}
which predicts two cases:

\textbf{\emph{1. When $\overline{\beta}_1=\overline{\beta}_3$}}

\h In this case, the first equation of (\ref{Eq29}) is satisfied
if $ \cos (2\alpha) = 1 $, where $ \alpha = m \pi $, and $m$ are
integers (note that the second equation of $(\ref{Eq29})$ is also satisfied). So

\begin{equation}
d=\frac{m}{\cos\theta_2}\frac{\lambda}{2n_2} \,\,,
\label{Eq30}
\end{equation}
where $\lambda = 2\pi c / \om$.

\textbf{\emph{2. When $\overline{\beta}_{1}\neq\overline{\beta}_{3}$}}

\h The second equation of (\ref{Eq29}) is satisfied if $ \sin(2
\alpha)=0$, so $ 2\alpha = m \pi $, with $ m $ an odd number,
otherwise the first equation can not be satisfied. So, the film
thickness is given by
\begin{equation}
d=\frac{m}{\cos\theta_2}\frac{\lambda}{4n_2}\,\, ,
\label{Eq33}
\end{equation}
where $\lambda = 2\pi c / \om$.
With respect to the first equation of (\ref{Eq29}), by knowing that $ \cos (2\alpha)=-1$, we have
\begin{equation}
\overline{\beta}_{2}=\sqrt{\overline{\beta}_1\overline{\beta}_3}
\label{Eq31}
\end{equation}
So, the refractive index of the film must be

\begin{equation}
n_2=\frac{\sqrt{n_1\cos\theta_1n_3\cos\theta_3}}{\cos\theta_2}
\label{Eq32}
\end{equation}

\h With this, we can ensure that de central Bessel beam of the
incident FW will be totally transmitted, and so we can expect that
the transmission of the entire FW beam will be maximized.

\h Now, let us test this with an example.

\textbf{Third example}

\h Let us assume the refractive indexes $n_1 = 1 $ (air) and
$n_3= 1.5$ (glass lens) for the first and third layer,
respectively. The anti-reflective film, deposited
over the glass surface in contact with air, will have refractive
index $n_2$ and thickness $ d $ (both to be calculated).

\h We can use the compensation method to obtain, in the third medium, a zero-order FW beam whose the desired longitudinal intensity pattern, $|F(z)|^2$, is given (let us suppose) by Eq.(\ref{F}), with $L=0.3$m,
$\delta=0.035$m, $l_1=0.03$m, $l_2=l_1+\delta$, $l_3=l_2+\delta$
and $l_4=l_3+\delta$ and a spot radius $r_0 = 20\mu$m. With the value of $r_0$ we obtain $Q = 0.99997 n_3\om /c$, from which we can calculate, for the anti-reflective film, $n_2 = 1.22$ and $d = 0.129\mu$m (for the first maximum transmission). With all the values of the refractive indexes and also of $d$, we can proceed by using the compensation method to obtain the incident beam that will result into the desired FW in the last medium.

\begin{figure}[h]
\subfloat[]{
    \includegraphics[height=5.5cm]{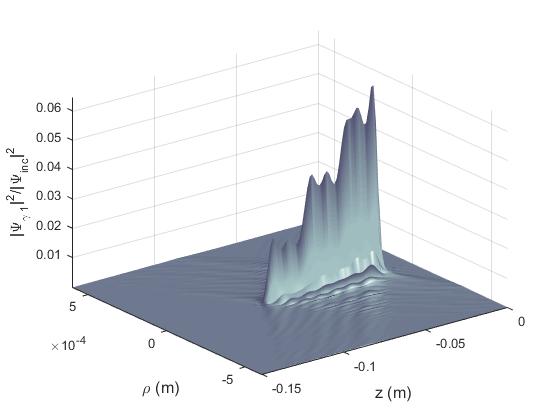}
}
\subfloat[]{
    \includegraphics[height=5.5cm]{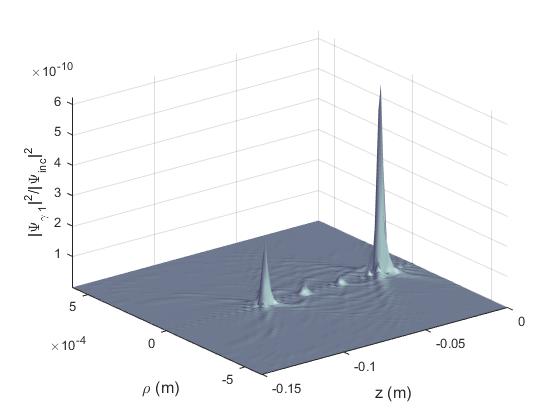}
}
\quad
\subfloat[]{
     \includegraphics[height=5.5cm]{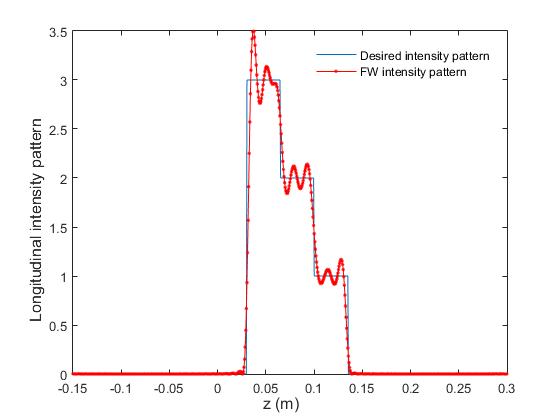}
}
\subfloat[]{
     \includegraphics[height=5.5cm]{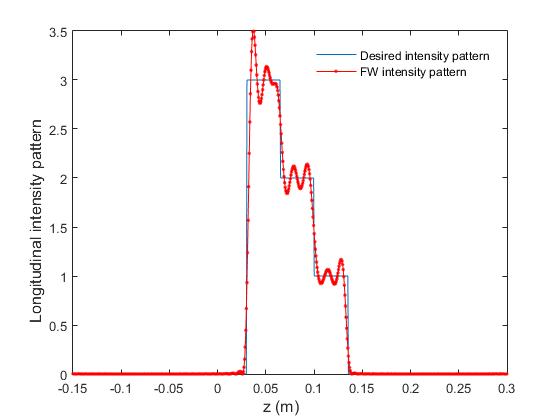}}
\caption{
 Intensity of the reflected FW in a setting (a) without the film and (b) with the film, both normalized with a peak intensity of the incident beam. Comparison between the intensity of the desired longitudinal pattern of the transmitted FW (c) without the film and (e) with the film. The solid line represents the function $F(z)$ and the dotted one the FW.} \label{gdimotes}
\end{figure}

%For the compensation method, we have, $Q=0.99997k$.
%
%\h Adopting a zero order incident FW $(\nu=0)$, let us suppose the
%desired longitudinal intensity pattern, $|F(z)|^2$, in the third
%(last) medium is given given by Eq.(\ref{F}), with $L=0.3$m,
%$\delta=0.035$m, $l_1=0.03$m, $l_2=l_1+\delta$, $l_3=l_2+\delta$
%and $l_4=l_3+\delta$ and that the desired spot radius is $r_0 =
%2\mu$m.

\h Figure 6 shows the effect of using (or not) the anti-reflective
film. We can notice that the use of the film almost eliminates the
reflection of the incident FW in the interface air-film, as expected. This fact implies
that with the implementation of the thin film a lower power is
necessary to generate the desired FW in the third medium.

%\newpage

\section{The transfer-matrix method}

\h In this section we are going to develop a transfer-matrix
formulation, which will enable us to speed up the implementation
of the longitudinal intensity modelling of diffraction resistant
beams in the last medium of a stratified dielectric structure.
More specifically, with this matrix method, once we have chosen
the desired FW for the last medium, we will be able to quickly
calculate the incident beam to be generated in the first medium in
such a way that the resulting beam in the last one is exactly
that we have previously chosen. Such a formulation will be very
suitable for stratified media with a large number of layers.

\h As we know, the incident FW beam is given by Eq.(\ref{inc}).
Now, let us to rewrite the Eqs.(\ref{psim}), which describe the
resulting waves within each medium, in the following compact form:

\bb \Psi_{m}(\rho,\phi,z) = \sum\limits_{q=-N}^{N} \psi_{m
q}(\rho,\phi,z) \,\,\,\, , \label{psim2} \ee with

\bb \psi_{m q}(\rho,\phi,z) = \Ncal_{\nu}A_q\tau_{mq} J_\nu(h_{
q}\rho)e^{i\nu\phi}e^{i\beta_{mq}z} + \Ncal_{\nu}A_q\Gamma_{mq}
J_\nu(h_{ q}\rho)e^{i\nu\phi}e^{-i\beta_{mq}z} \,\,\, ,
\label{psimq} \ee for $m = 1,2,..,M$.

\h From Eq.(\ref{psimq}), it is not difficult to show the
following matrix equation, for $m=2,..,M-1$, relating $\psi_{m q}$
and its derivative $\pa_z \psi_{m q}$ evaluated on the interface
at $z=d_{m}$ with their values on the preceding interface
$z=d_{m-1}$ :

\bb  \left[ \begin{array}{cc}
 \psi_{m q}(\rho,\phi,z=d_m) \\
\pa_z\psi_{m q}(\rho,\phi,z=d_m)  \end{array} \right] \ug
\texttt{\textbf{M}}_m^q \,\left[
\begin{array}{cc}
 \psi_{m q}(\rho,\phi,z=d_{m-1}) \\
\pa_z\psi_{m q}(\rho,\phi,z=d_{m-1})  \end{array} \right] \,\,,
\label{eqmatrix1} \ee
with

\bb \texttt{\textbf{M}}_m^q \ug \left[ \begin{array}{cc}
 \cos(\beta_{m q}\Delta_m) & \frac{1}{\beta_{m q}}\sin(\beta_{m q}\Delta_m) \\
-\beta_{m q}\sin(\beta_{m q}\Delta_m) &\cos(\beta_{m q}\Delta_m)
\end{array} \right] \,\, ,  \label{Mmq} \ee
and

\bb \Delta_m \ug d_{m} - d_{m-1} \label{D} \ee

\h Now, from Eqs.(\ref{eqmatrix1} - \ref{D}) and due to the boundary conditions asserting the continuity of
$\psi_{m q}$ and $\pa_z \psi_{m q}$ through each plane interface at
$z=d_m$ ($m=1,..,M-1$), we can write down the following equation
relating $\psi_{M q}$ and $\pa_z \psi_{M q}$ with $\psi_{1 q}$ and
$\pa_z \psi_{1 q}$:

\bb  \left[ \begin{array}{cc}
 \psi_{M q}(\rho,\phi,z=d_{M-1}) \\
\pa_z\psi_{M q}(\rho,\phi,z=d_{M-1})  \end{array} \right] \ug
\texttt{\textbf{M}}^q \,\left[
\begin{array}{cc}
 \psi_{1 q}(\rho,\phi,z=0) \\
\pa_z\psi_{1 q}(\rho,\phi,z=0)  \end{array} \right] \,\,,
\label{eqmatrix2} \ee
 with the transfer-matrix,

\bb \texttt{\textbf{M}}^q \ug \left[ \begin{array}{cc}
 \texttt{\textbf{M}}_{11}^q & \texttt{\textbf{M}}_{12}^q \\
\texttt{\textbf{M}}_{21}^q & \texttt{\textbf{M}}_{22}^q
\end{array} \right] \,\, , \label{Mq} \ee
given by

\bb \texttt{\textbf{M}}^q=\texttt{\textbf{M}}^q_{M-1}\cdot...\cdot
\texttt{\textbf{M}}^q_3\cdot \texttt{\textbf{M}}^q_2   \ee

\h From Eqs.(\ref{psimq},\ref{eqmatrix2},\ref{Mq}) we get

\begin{equation}
\Gamma_{1q}=\frac{(\texttt{\textbf{M}}^q_{21}+\beta_{1 q}\beta_{M
q}\texttt{\textbf{M}}^q_{12})+i(\beta_{1
q}\texttt{\textbf{M}}^q_{22}-\beta_{M
q}\texttt{\textbf{M}}^q_{11})}{(-\texttt{\textbf{M}}^q_{21}+\beta_{1
q}\beta_{M
q}\texttt{\textbf{M}}^q_{12})+i(\beta_{1q}\texttt{\textbf{M}}^q_{22}+\beta_{M
q}\texttt{\textbf{M}}^q_{11})} \label{gama1q}
\end{equation}

and

\begin{equation}
\tau_{M q}=2i\beta_{1 q}e^{-i\beta_{M
q}d_{M-1}}\left[\frac{1}{-\texttt{\textbf{M}}^q_{21}+\beta_{1 q}\beta_{M q}
\texttt{\textbf{M}}^q_{12}+i(\beta_{1
q}\texttt{\textbf{M}}^q_{22}+\beta_{M
q}\texttt{\textbf{M}}^q_{11})}\right] \label{tauMq}
\end{equation}

\h Now, with respect our compensation method, we can see from the receipt of four steps developed in Section 4 that Eq.(\ref{tauMq}) can be directly used in Eq.(\ref{Aq2}) to furnish the coefficients of the Bessel beam superposition (\ref{inc}) which defines the incident beam (in the first medium) that, after crossing the stratified structure, will result into the desired FW beam in the last medium.

\h Naturally, we can also use Eq.(\ref{gama1q}) to get the reflected beam in the first medium.

\h As we have already said, this matrix-transfer formulation can be very effective in the application of the compensation method in cases where the stratified medium possesses a large number of layers. It can be equally useful for applying the compensation method to a continuously inhomogeneous medium, with refractive index $n = n(z)$, if we discretize it as a sequence of thin layers, each one with its (different) constant refractive index.

\section{Conclusion}

\h In this paper, we describe analytically the propagation of Frozen-Wave-type
beams through non-absorbing stratified media, showing how the desired beam pattern for the last medium
is affected by the inhomogeneities. More importantly, we develop a novel method for
constructing an incident FW beam capable of compensating such
inhomogeneity effects, so rendering the desired
spatially-shaped-diffraction-resistant-beam in the last material
medium. In this context, we also develop a matrix method to deal
with stratified media with large number of layers.

\h Additionally, we undertake some discussion about minimizing reflection of the incident FW beam on the first material interface by using thin films.

\h The results here presented can have important applications in remote sensing, medical and militar purposes, non invasive optical measurements, etc..

%\section*{References}
%
%\bibliography{artigobib}

\end{document}